\documentclass[12pt,prb,aps,,showpacs,preprint,nofootinbib]{revtex4}

\usepackage{graphicx}
\usepackage{float}
\usepackage{amsmath}
\usepackage{amsfonts}

\begin{document}

\title{Toward the Jamming Threshold of Sphere Packings: Tunneled Crystals}

\author{S. Torquato}

\email{torquato@electron.princeton.edu}

\affiliation{\emph{Department of Chemistry}, \emph{Princeton University}, Princeton
NJ 08544}

\affiliation{\emph{Program in Applied and Computational Mathematics}, \emph{Princeton
University}, Princeton NJ 08544}

\affiliation{\emph{Princeton Institute for the Science and Technology of Materials, Princeton University}, Princeton NJ 08544}

\affiliation{\emph{Princeton Center for Theoretical Physics, Princeton University}, Princeton
NJ 08544}

\author{F. H. Stillinger}

\affiliation{\emph{Department of Chemistry}, \emph{Princeton University}, Princeton
NJ 08544}

\begin{abstract}

We have discovered a new family of three-dimensional crystal sphere packings 
that are strictly jammed (i.e., mechanically stable) and  yet possess an anomalously low density.  
This family constitutes an uncountably infinite number of crystal packings that are
subpackings of the densest crystal packings and are characterized
by a high concentration of self-avoiding ``tunnels" (chains of vacancies) 
that permeate the structures. The fundamental geometric characteristics
of these tunneled crystals command interest in their own
right and are described here in some detail. These include the lattice vectors
(that specify the packing configurations),
coordination structure, Voronoi cells, and density fluctuations.
The tunneled crystals are not only candidate structures for achieving the jamming threshold
(lowest-density rigid packing), but may have substantially broader significance for 
condensed matter physics and materials science.

\end{abstract}

\maketitle

\section{Introduction}

Hard-particle models have played a substantial and insightful role in the historical 
development of statistical mechanics.  In particular, this is true
for the venerable hard-sphere model in $d$-dimensional
Euclidean space $\mathbb{R}^d$ in which the hyperspheres only interact
with an infinite repulsion for overlapping configurations.
Hard-sphere packings have provided a rich source of outstanding theoretical problems and have served
as useful starting points to model the structure of granular media,\cite{Ed01} liquids, \cite{Ha86,To02a} glasses, \cite{To02a}
 crystals, \cite{Chaik95} living
cells, \cite{To02a} and random media. \cite{To02a} Sphere packing problems
have inspired scientists and mathematicians at least since the time of Kepler
and continue to present open challenging problems. \cite{Co98,As00,Co03,Ha05,To06b}

One of the perennially popular aspects of hard-sphere many-body systems concerns
their ``jamming" properties, i.e., their mechanically stable packing arrangements.  Jamming
behavior of sphere packings is intimately related to classical ground-state structures
and to glassy states of matter. The present paper 
concentrates on one portion of that packing arrangement issue that to the best of our 
knowledge has not previously been explored, namely, the ``strict" jamming threshold 
of three-dimensional sphere packings.

Three broad and mathematically precise
``jamming" categories of sphere packings can be distinguished depending on the nature of their 
mechanical stability; \cite{To01} see also Ref. \onlinecite{To03a}.  In order of increasing stringency (stability) for
a finite system of hard spheres, these are the following:
\begin{description}
\item [Local~jamming:]Each particle in the system is locally trapped by
its neighbors, i.e., it cannot be translated while fixing the positions
of all other particles. 
\item [Collective~jamming:]Any locally jammed configuration is collectively jammed if no
subset of particles can simultaneously be displaced so that its members
move out of contact with one another and with the remainder set. An
equivalent definition is to ask that all finite subsets of particles
be trapped by their neighbors.
\item [Strict~jamming:]Any collectively jammed configuration that disallows
all globally uniform volume-nonincreasing deformations of the system
boundary is strictly jammed.
\end{description}
It is important to note that the jamming category depends on the
boundary conditions employed. For example, hard-wall boundary
conditions \cite{To01} generally yield different
jamming classifications from periodic boundary conditions. \cite{Do04}
These jamming classifications are closely related to the concepts of ``rigid" and ``stable"
packings found in the mathematics literature. \cite{Con98}
Rigorous and efficient linear-programming algorithms have been devised
to assess whether a particular hard-sphere configuration is locally, collectively, 
or strictly jammed. \cite{Do04,Do04b}

\begin{figure}[H]
\centerline{\includegraphics[height=3.0in,keepaspectratio]{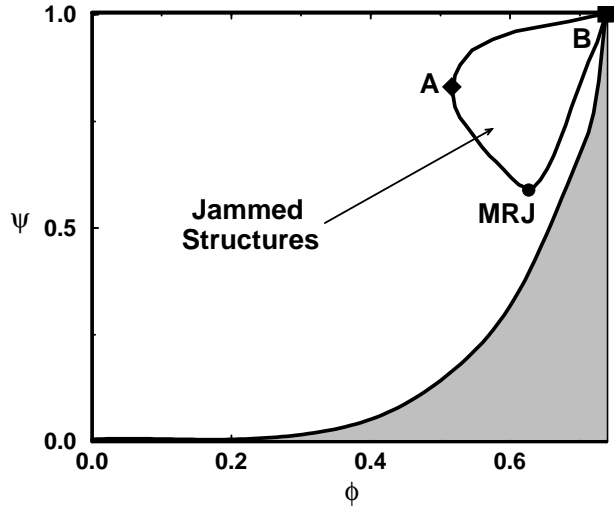}}
\caption{ A highly schematic plot of the jammed subspace in the density-disorder
($\phi$-$\psi$) plane, taken from Ref. \onlinecite{To00}. Point A corresponds to the lowest-density jammed
packing, and it is intuitive to expect that a certain ordering will be
needed to produce low-density jammed packings. Point B corresponds
to the densest jammed packing. Point MRJ represents
the maximally random jammed state, i.e., the most disordered
state subject to the jamming constraint.}
\label{phi-psi}
\end{figure}

The enumeration and classification of both ordered and disordered jammed circular disk and 
sphere packings for the various jamming categories is an outstanding problem.
Since one cannot enumerate all possible packings even for a small number of particles, it is 
desirable to devise a small set of parameters that can characterize packings well. One 
important property of a sphere packing is the packing fraction $\phi$, which is defined 
to be the fraction of space covered by the particles. Another useful
way to characterize a packing  is via scalar order metrics. \cite{To00,To02a}
An order metric $\psi$ is a well-defined scalar function of
a configuration of spheres and is subject to the normalization $0\le \psi \le 1$. For any
two states $X$ and $Y$, $\psi(X) > \psi(Y)$ implies that state $X$ is to be
considered as more ordered than state $Y$. Candidates for such an order metric
include various translational and orientational order
parameters \cite{To00,To02a,Ka02b} but the search for better order metrics is still
very active. Figure \ref{phi-psi} from Ref. \onlinecite{To00} shows a highly schematic region of
feasible hard-sphere packings in the $\phi$-$\psi$ plane, which has been called
an ``order map." It is clear that only a small subset of this feasible region will be occupied
by jammed packings for a given jamming category, as indicated in Fig. \ref{phi-psi}. 
The following extremal points in the jammed region are particularly interesting:

\begin{enumerate}
\item Point A corresponds to the lowest-density jammed
packing, i.e., the jamming threshold, and its location strongly depends on the jamming
category used. We denote by $\phi_{\mbox{\scriptsize min}}$ the corresponding
jamming-threshold packing fraction, which is expected to be characterized
by a high degree of order, as discussed in more detail below., As discussed below,
local jamming is a very weak condition compared to
collective or strict jamming.

\item Point B corresponds to the most dense jammed packing.
It has of course already been identified to be a triangular
lattice packing for circular disks, and the face-centered cubic lattice
packing and its stacking variants for spheres. But much less is known about polydisperse
packings \cite{Fe64,Ka02a,Uc04} or packings of nonspherical particles. \cite{Do04c,Co06}

\item The  MRJ point represents the {\it maximally random jammed}
state, \cite{To00} which has been suggested to replace the ill-defined
random close packed (RCP) state. \cite{rcp}\ The MRJ state is the most disordered
jammed packing in a given jamming category (i.e., locally, collectively,
or strictly jammed). The MRJ state is well-defined
for a given jamming category and choice of order metric.
The strict MRJ state can be regarded to be the {\it prototypical} glass --
it is the most disordered packing arrangement that is
able to withstand shear forces. 
\end{enumerate}
It is crucial to note that the {\it order map} shown in Fig. \ref{phi-psi}
is independent of the protocol used to generate a hard-sphere configuration.
In practice, one can use a variety of protocols to produce jammed configurations
in order to delineate the boundary of the jammed region shown
in Fig. \ref{phi-psi}, as was partially done in Ref. \onlinecite{Ka02b}.
Moreover, the frequency of occurrence of a particular configuration is
irrelevant in so far as the order map is concerned. 
In other words, the order map emphasizes a  statistical-geometric approach 
to packing by characterizing single configurations
regardless of their occurrence probability. Therefore, ensemble
methods  that inherently produce ``most probable" configurations
might miss interesting extremal points in the order
map, such as point A.

The preponderance of work on sphere packings has been devoted to the
determination of point B in Fig. \ref{phi-psi} for low as well as high
dimensions. \cite{Ha05,Co98}
It is known that the densest arrangements of monodisperse disks in two dimensions
and spheres in three dimensions are strictly jammed. \cite{To01,Do04}
This implies that shear moduli of these packings are infinitely
large. The densest sphere packings have a packing fraction given  by
\begin{equation}
\phi_{\mbox{\scriptsize max}}= \frac{\pi}{\sqrt{18}}= 0.74048\ldots
\label{max}
\end{equation}
It has long been known empirically that this maximum is attained both by the 
face-centered cubic (fcc) and the hexagonal close packed (hcp) crystal structures, 
as well as by their stacking hybrids.  A mathematically rigorous proof of Eq. (\ref{max}) 
has only recently appeared. \cite{Ha05}

We are interested in  characterizing point A in Fig. \ref{phi-psi} for collectively
and strictly jammed packings, which has received far less attention
than the determination of point B. Specifically, it is desired to identify 
such packing arrangements and the corresponding jamming-threshold
packing fraction $\phi_{\mbox{\scriptsize min}}$. It is possible to arrange hard spheres in space, 
subject only to the weak locally jammed criterion, so that the resulting packing fraction  
is arbitrarily close to zero. \cite{Bo64,St03}  But demanding either collective jamming or 
strict jamming evidently forces $\phi$ to equal or exceed a lower limit $\phi_{\mbox{\scriptsize min}}$ that is well above zero.  
No rigorous theory or even empirical study has heretofore convincingly determined $\phi_{\mbox{\scriptsize min}}$ for collectively or 
strictly jammed monodisperse hard spheres.

The present paper is devoted to a description of a class of strictly jammed 
three-dimensional sphere packings with anomalously low 
packing fraction, substantially lower indeed than the lowest previously known result. \cite{Ka02b,St03} 
This class appears to be a new family of crystal structures.
We do not know if the resulting packing fraction for this class actually attains $\phi_{\mbox{\scriptsize min}}$.  
The packing  structures involved rely on an earlier observation that linear arrays, or tunnels, of vacancies 
generated in close-packed crystals, even with branching, do not destroy the mechanical stability 
of the resulting structures. \cite{St03}

The following Section II provides some background on the problem
and describes previous work on low-density jammed packings.
Section III  provides a structural characterization of the class of low-$\phi$ tunneled structures.  Via application of 
a computational test, \cite{Do04,Do04b} we establish that strict jamming is attained in these structures.  
The final Section IV contains discussion, including the informal argument that approaching or 
attaining $\phi_{\mbox{\scriptsize min}}$ requires a regular periodic structure, not an amorphous packing.

\section{Background and Previous Work}

One way to reduce  the density of a strictly jammed packing
while retaining the strict jamming characteristic is  to selectively remove  
subsets of spheres from the fcc, hcp, or hybrid close packed crystals.  This leaves behind 
an array of vacancies.  In this approach it is important to avoid removing triads of spheres 
that were in mutual contact in the starting crystal, i.e., a compact equilateral triangle of spheres, 
because that leads to local instability. \cite{St03}  However, one viable option involves removing 
one-quarter of the spheres from an fcc crystal, specifically those that 
constitute one of its four simple-cubic sublattices.  The result remains strictly jammed \cite{St03} 
and thus leads to the following bound:
\begin{equation}
\phi_{\mbox{\scriptsize min}} \le \frac{\pi}{2^{5/2}} = 0.55536\ldots
\end{equation}

In addition to the vacancy-containing crystals just described, collectively and 
strictly jammed packings with $\phi < \phi_{\mbox{\scriptsize max}}$ also exist with 
irregular (non-periodic) sphere arrangements. Indeed, in Ref. \onlinecite{To00}
it was shown that using the Lubachevsky-Stillinger algorithm, \cite{Lu90} one can
produce such packings in non-trivial range of packing fraction:
\begin{equation}
0.64 < \phi < 0.74048\ldots
\end{equation}                                                                
where 0.64 corresponds to the packing fraction of the MRJ state.
In fact, we have conjectured that the Lubachevsky-Stillinger
packing algorithm  typically produces packings along the
right (maximally dense) branch from the MRJ point to the maximally
dense point B in Fig. \ref{phi-psi}. \cite{Do05} Importantly, we do not know of an algorithm
that can systematically produce packings along the left (minimally dense)
branch without relying on some random removal process. Indeed,
it has been shown that by randomly diluting the
fcc packing (subject to the constraint that no compact
equilateral triangular vacancies are created), strictly jammed
packings with a packing fraction of 0.52 can be created. \cite{Ka02b}
It should not escape notice that these considerations 
suggest that amorphous sphere configurations cannot attain 
the jamming threshold $\phi_{\mbox{\scriptsize min}}$. Rather
it is very plausible that $\phi_{\mbox{\scriptsize min}}$
is achieved by structures characterized by a large order metric $\psi$ value,
as schematically indicated in Fig. \ref{phi-psi}. In other words,
 $\phi_{\mbox{\scriptsize min}}$ is likely to be realized by crystal (i.e., periodic) packings. \cite{order}
Typical large, jammed packings produced via experimental as well as computer-simulation protocols
are characterized by a significant degree of disorder, and therefore such protocols would
never find such crystal candidates because they are sets of zero measure.

{\it Isostatic} packings are jammed packings that have the minimal
number of contacts to maintain a particular jamming classification,
a situation that are normally associated with amorphous
packing such as the MRJ state (c.f. Fig. \ref{phi-psi}).
In the limit of an infinitely large packing, collective
and strict jamming become equivalent constraints and the corresponding
isostatic condition implies an average of $2d$ contacts per particle, \cite{Do05}
where we recall that $d$ is the space dimension. Thus, ordered but strictly jammed 
sphere packings in $\mathbb{R}^d$ with $2d$ contacts per particle
would seem to be natural candidates to achieve the jamming threshold.
Indeed, in $\mathbb{R}^2$, the so-called ``reinforced" Kagom{\' e}
packing with precisely 4 contacts per particle is evidently 
the lowest density strictly jammed subpacking of the triangular lattice
packing \cite{Do04,Do06} with $\phi_{\mbox{\scriptsize min}} = \sqrt{3} \pi/8 = .68017\ldots$. 
The $d$-dimensional generalization of the two-dimensional
Kagom{\' e} packing has exactly $2d$ contacts per particle
because each particle is the vertex of vertex-sharing
simplices \footnote{The $d$-dimensional Kagom{\' e} packing 
contains $d+1$ spheres per fundamental cell i.e., it
has a $(d+1)$-particle basis. The  centroids of the simplices
of this structure are the sites of the $d$-dimensional {\it diamond} crystal
that possesses a 2-particle basis and placing the largest
nonoverlapping hypersphere at each of these sites produces the  densest $d$-dimensional 
diamond packing.  The ``two-dimensional diamond" packing is nothing more than the ``honeycomb"
packing, which is the basic building block used to create
the tunneled three-dimensional crystals that are the focus of this paper. Placing 
the largest nonoverlapping hypersphere at each of the midpoints
of the ``bonds" joining the sites of the
$d$-dimensional diamond packing yields the densest $d$-dimensional Kagom{\' e} packing.
Detailed geometrical characteristics
of the latter packing for arbitrary $d$ will be reported in a future study.}
and would appear to achieve the desired 
jamming threshold. In three dimensions, this structure
is the well-known pyrochlore crystal that has received considerable attention 
because such material structures can exhibit exotic magnetic behavior; see Refs. \onlinecite{An56}
and \onlinecite{Ca99}, and references therein. The three-dimensional
Kagom{\' e} packing possesses a rather low
packing fraction ($\phi = \pi/\sqrt{72}= 0.37024\ldots$), but,
unfortunately, it contains equilateral-triangle-vacancy cluster
and therefore cannot be collectively or strictly jammed.
Thus, the $d$-dimensional Kagom{\' e} packing  is
not strictly jammed for $d \ge 3$.

Note that in a Bravais lattice packing, the space $\mathbb{R}^d$ can be geometrically divided into identical
regions called {\it fundamental cells}, each of which contains the center
of just one sphere. Non-Bravais-lattice packings include periodic packings, in which 
there is more than one sphere per fundamental cell, as well as disordered
packings.

\section{Tunneled Close-Packed Sphere Packings}

     We begin by reminding the reader about the elementary distinctions between the fcc, 
the hcp, and the hybrid close-packed structures.  All can be conveniently viewed as stacks of planar 
triangular arrays of spheres, within which each sphere contacts six neighbors.  These triangular 
layers can be stacked on one another, fitting spheres of one layer into "pockets" formed by 
nearest-neighbor triangles in the layer below.  At each such layer addition there are two choices of 
which set of pockets in the layer below are to be filled.  A lower layer with lateral position to be 
called A, is then surmounted with the next layer in lateral position B or C.  A third layer subsequently 
can revert to lateral position A, or can be C on a second layer B, or B on second layer C.  
The fcc structure is a Bravais-lattice packing that  utilizes the repeating pattern:
\begin{equation}
\ldots\; \mbox{ABCABCABC} \; \ldots 
\label{fcc}  
\end{equation}
while the hcp case (a periodic non-Bravais-lattice packing) corresponds to:
\begin{equation}
\ldots\; \mbox{ABABABAB} \; \ldots  
\label{hcp}                                         
\end{equation}
Hybrid close-packed structures utilize other A,B,C patterns of lateral positions, never 
immediately repeating one of these three letters. Since there are two ways to 
place each layer after the second, there is an uncountable infinity of distinct 
packing schemes, all with the same density. These are called the Barlow packings 
\cite{Ba83} and include {\it random} stacking variants (i.e., the two ways to place each layer 
after the second occur with equal probabilities.). In the latter case, 
there is no repeating pattern, as exhibited by the following partial sequence:
\begin{equation}
\ldots\;{ABACBACBCA}\;\ldots
\end{equation}

\begin{figure}[bthp]
\centerline{\includegraphics[height=2.0in,keepaspectratio]{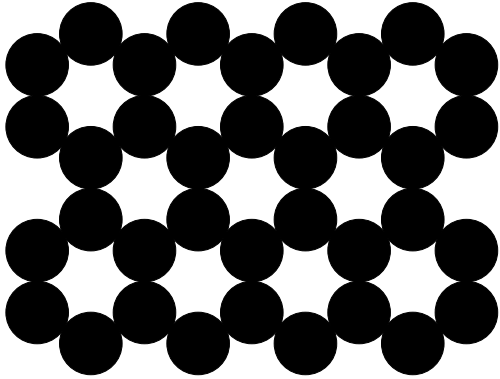}}
\caption{ A portion of a honeycomb layer structure.}
\label{honey}
\end{figure}

The tunneled crystal structures to which this paper is devoted can similarly be classified 
by a layer lateral displacement A,B,C code.  However, the constituent planar layers 
to be stacked are not triangular, but have the lower density ``honeycomb" pattern. 
The latter is illustrated in Fig. \ref{honey}. This amounts to the preceding triangular 
layer with one-third of its spheres removed in a periodic pattern.  
Each remaining sphere in the honeycomb layer contacts three neighbors in that layer.  
A periodic primitive or fundamental cell for the honeycomb structure in two dimensions 
contains two spheres, not just one as for the triangular-lattice
 layer.

By adapting the three jamming category definitions to two dimensions, one immediately 
discovers that the honeycomb structure by itself is only locally jammed. \cite{To01}  It is 
easy to see from Fig. \ref{honey} that the set of six particles surrounding any vacancy 
can be rotated as a unit about its center to eliminate six contacts within the layer. 
Repetition of this process, along with other subsequently allowed displacements, would 
totally un-jam the layer. However, this intra-layer instability is eliminated when 
honeycomb layers are placed one upon another, following one of the previously-described 
A,B,C codes. Note that there are three different choices for the direction of
the tunnels at each stacking stage. There are an uncountably infinite number
of such stacking arrangements that we refer to as the ``tunneled" crystals.

\subsection{Tunneled FCC Crystal}

\begin{figure}[bthp]
\centerline{\includegraphics[height=2.7in,keepaspectratio]{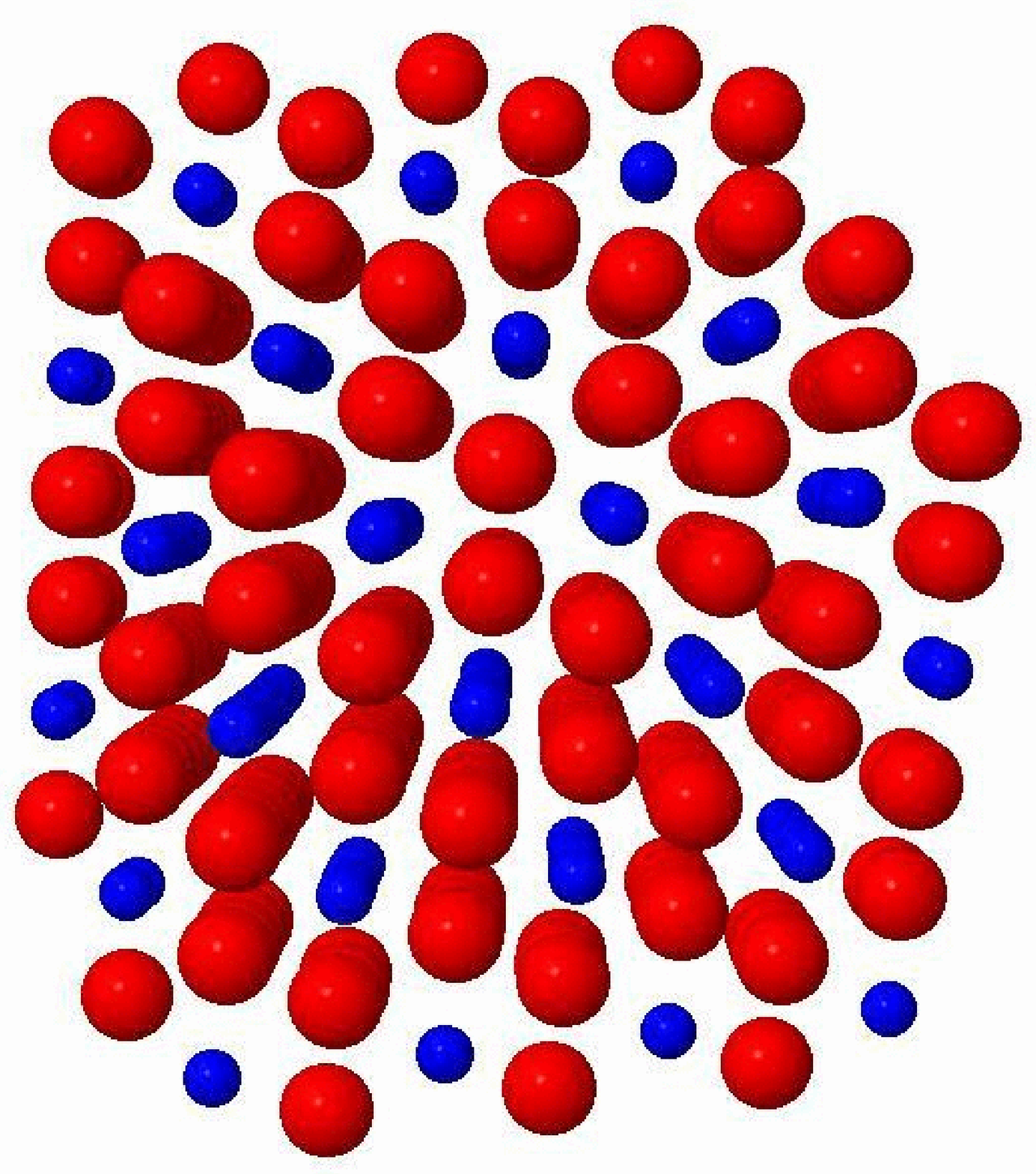}
\includegraphics[height=1.8in,keepaspectratio]{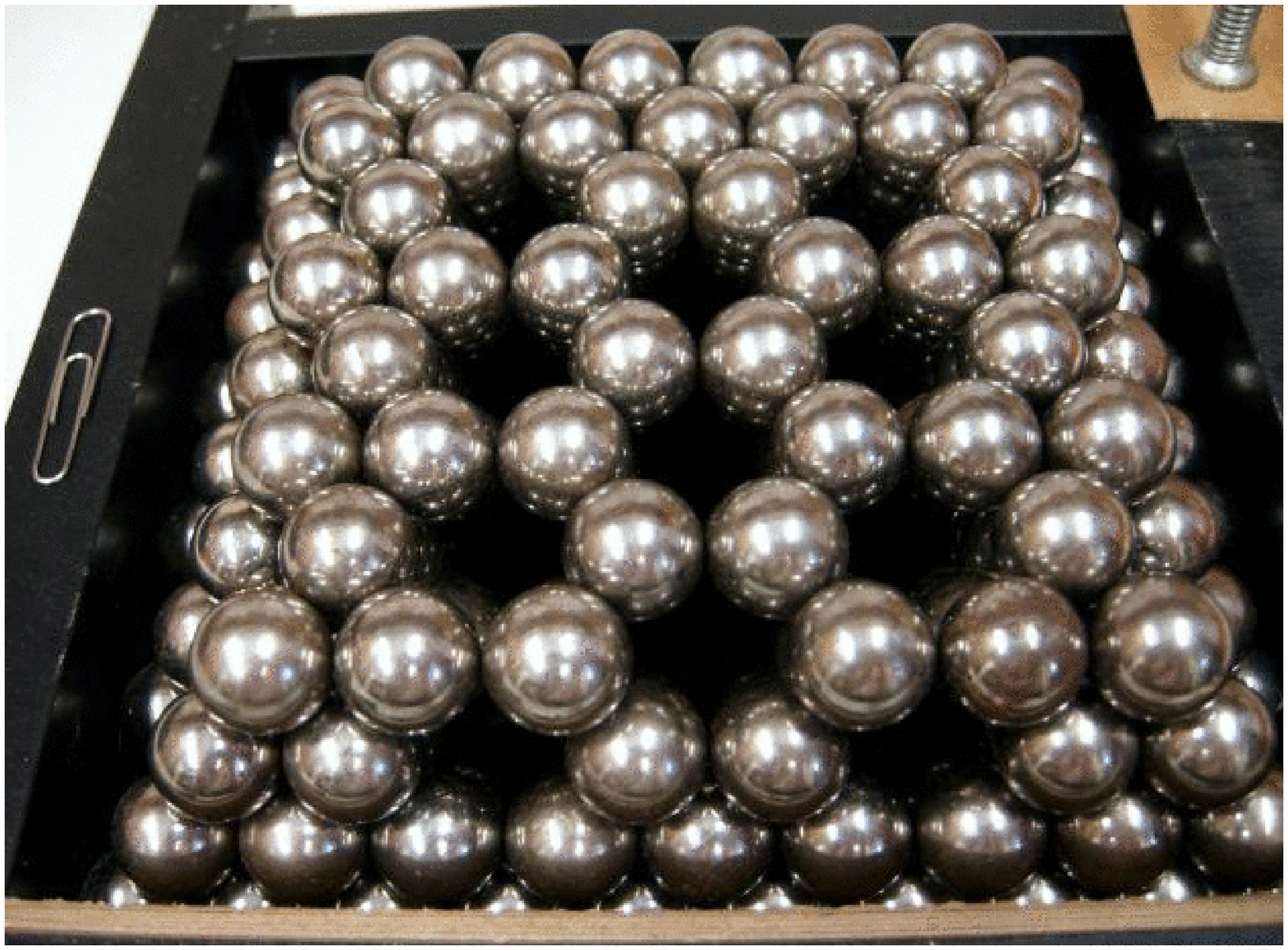}}
\caption{ Left panel: A view of the ``tunneled" fcc crystal looking
along the axis perpendicular to the honeycomb layers.
The vacancies are shown as smaller red particles and the actual particles are 
colored blue. The ``tunnels" consist of linear chains of vacancies and are parallel to
one another. Right panel: A photgraph of the tunneled fcc crystal built
up from ball bearings of diameter $5/8$ inches. This actual
construction shows the manifest stability of the tunnled fcc crystal.}
\label{fcc-tunnel}
\end{figure}

Consider first the fcc code (\ref{fcc}), which is depicted in the left 
panel of Fig. \ref{fcc-tunnel}. The right panel is a photograph of  a corresponding 
ball-bearing construction, which shows the tanifest stability of the packing.
The periodic result in three dimensions has a fundamental cell containing two 
spheres.  Assuming that the spheres have unit diameter, the basis vectors locating 
sphere centers for the fundamental cell can be assigned as follows:
\begin{equation}
{\bf a}_1= \sqrt{3} {\bf i}, \quad  {\bf a}_2= -\frac{\sqrt{3}}{2} {\bf i}  + \frac{3}{2} {\bf j}, \quad 
{\bf a}_3= -\frac{\sqrt{3}}{6} \,{\bf i}+ \frac{1}{2}\, {\bf j}+ \frac{\sqrt{2}}{3}\, {\bf k}.
\label{fcc-a}
\end{equation}
The additional sphere in this fundamental cell is located at
\begin{equation}
{\bf b_1}={\bf j}.
\label{fcc-b}
\end{equation}
As a result of using honeycomb layers to form this structure, the packing fraction is
\begin{equation}
\phi=\frac{2}{3}\phi_{\mbox{\scriptsize max}}=\frac{\sqrt{2}\pi}{9} = 0.49365\ldots
\label{phi-tunnel}
\end{equation}

\begin{figure}[H]
{\includegraphics[height=2.5in,keepaspectratio,clip=]{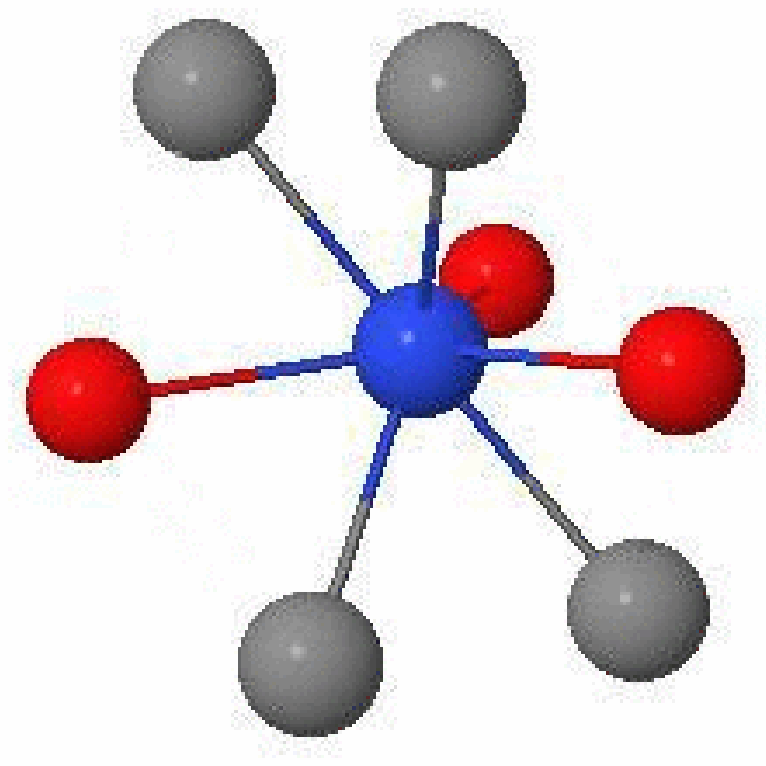}}
{\includegraphics[height=2.5in,keepaspectratio,clip=]{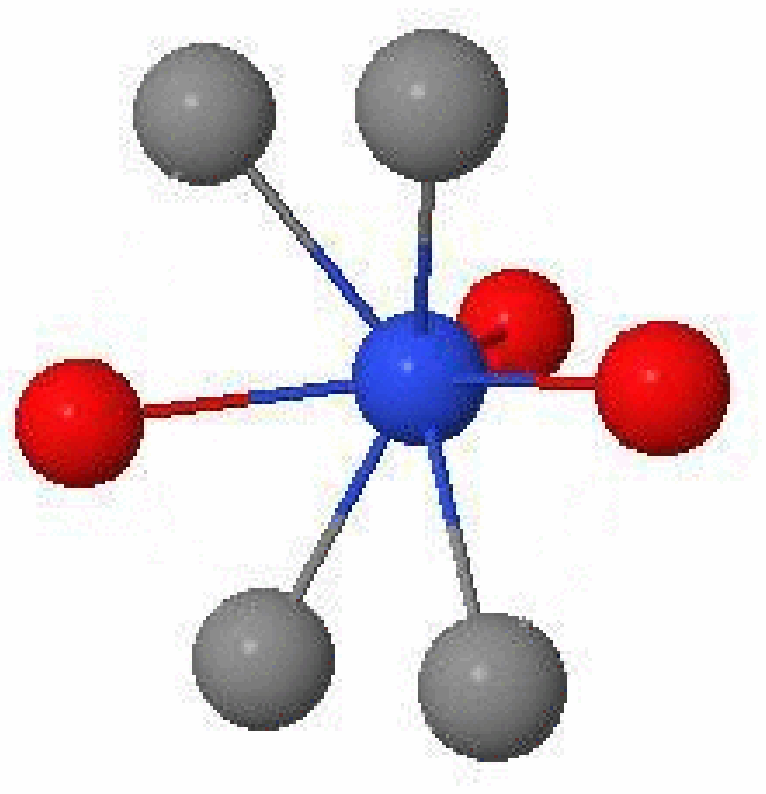}}
\caption{ The chiral pairs of seven contacting neighbor arrangements
in the tunneled fcc packing. 
The orientation of the three red spheres that contact a central blue
sphere within each honeycomb layer is the same in both chiral alternatives. }
\label{fcc-chiral}
\end{figure}

     Examination of the tunneled fcc crystal structure reveals that it contains a 
periodic array of parallel linear tunnels.  The direction of these tunnels is that 
of cube-face diagonals for the parent fcc crystal.  By symmetry this tunnel array 
could have been oriented in any one of six equivalent directions.  Each remaining 
sphere in the tunneled fcc crystal lies immediately next to three  tunnels. 
The number of neighbor contacts experienced by each sphere is seven, comprising 
three within its own honeycomb layer, and two each from the honeycomb layers 
immediately below and above.  The spatial arrangement of these seven contacting 
neighbors is chiral (i.e., the mirror image of one is not superimposable on 
the other), with equal numbers of left- and right-handed versions present 
(see Fig. \ref{fcc-chiral}). Clearly, the tunneled fcc crystal
has a lower symmetry than its parent unvacated fcc packing. 
 
\begin{figure}[bthp]
{\includegraphics[height=2.5in,keepaspectratio,clip=]{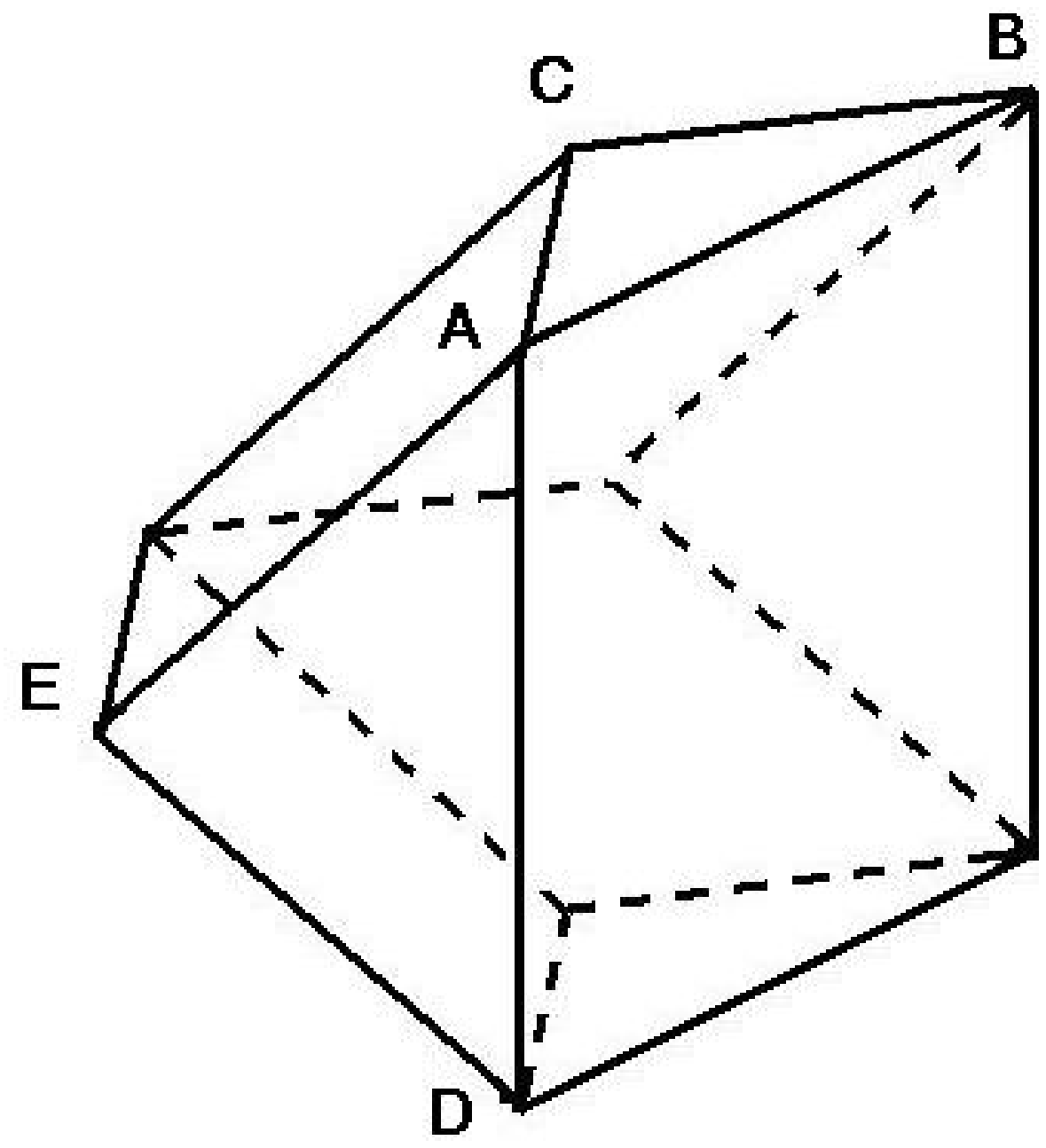}}\hspace{0.55in}
{\includegraphics[height=2.5in,keepaspectratio,clip=]{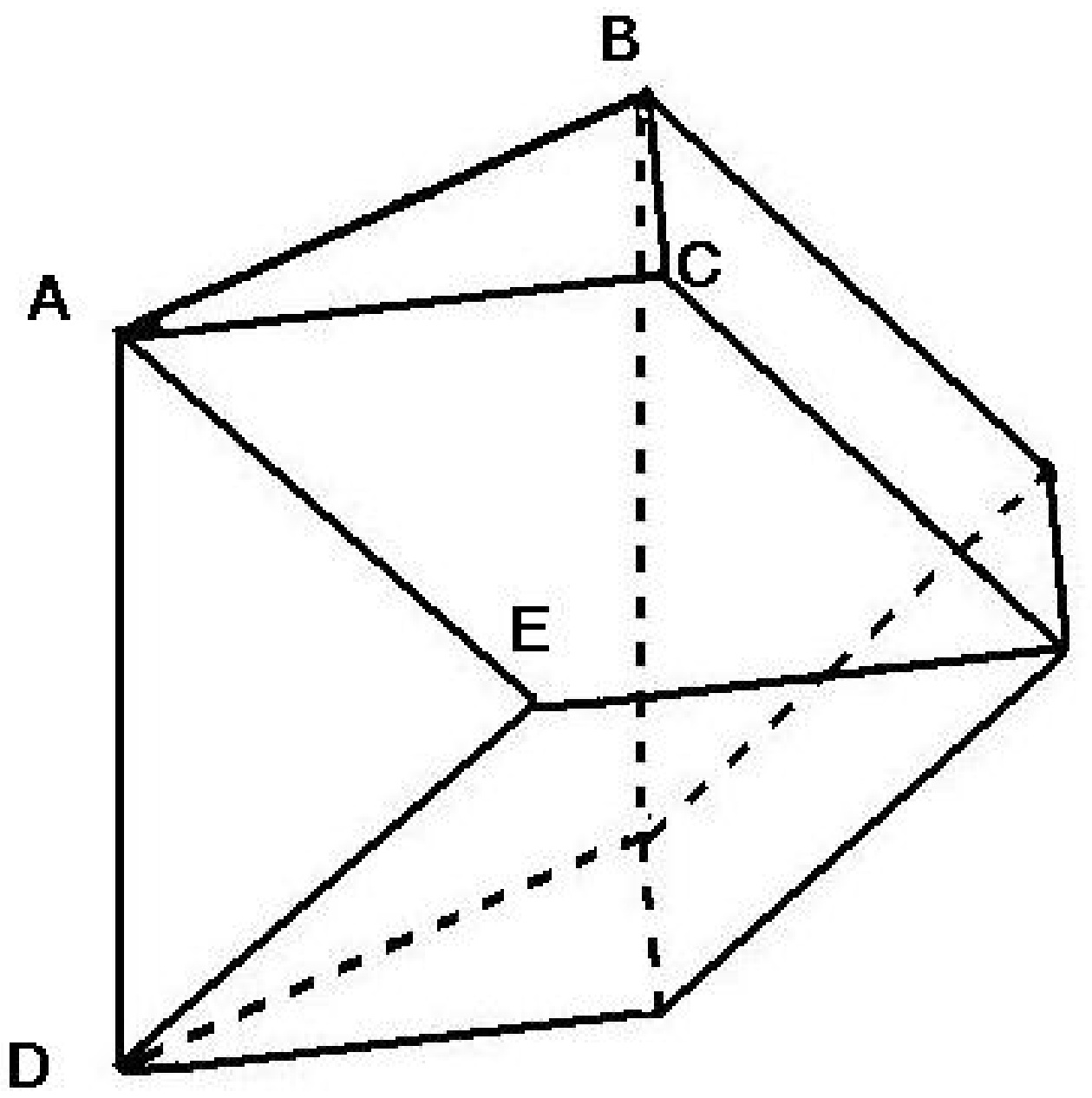}}
\caption{ The chiral pairs of Voronoi cells in the tunneled fcc packing. 
Each cell has 9 faces:
one rectangular face (with side lengths of 1 and 3/2), two isosceles  triangular faces
(with side lengths 1 and $1/\sqrt{2}$), two other isosceles  triangular faces
(with side lengths 3/2 and $1/\sqrt{2}$, and four rhombical faces (with side lengths 3/2 and $1/\sqrt{2}$). 
The edge lengths indicated are given as follows: $\overline{AB}=1$, $\overline{AC}=1/\sqrt{2}$, 
$\overline{BC}=1/\sqrt{2}$, $\overline{AD}=3/2$, $\overline{AE}=1/\sqrt{2}$,
and $\overline{DE}=1/\sqrt{2}$.
Each cell has 16 edges and 9 vertices. Merging the two rectangular faces 
(mirror planes) so that they coincide
results in a dodecahedron (12-faced polyhedron) that tiles the space.}
\label{fcc-voronoi}
\end{figure}

Associated with each sphere center is its {\em Voronoi cell}, 
which is defined to be the region of space nearer
to this center than to any other sphere center.
The Voronoi cells for any general point process are convex polyhedra whose interiors
are disjoint, but share common
faces, and therefore the union of all of the polyhedra
tiles the space. Not surprisingly, there are two types
of Voronoi cells for the tunneled fcc packing, one being 
the mirror image of the other (see Fig. \ref{fcc-voronoi})
and therefore, since these cells cannot be superimposed
on one another, they are chiral pairs. The volume of each Voronoi
cell is 3/2 times the volume of the Voronoi cell (rhombic
dodecahedron) of an unvacated fcc packing.

The vector ${\bf R} = n_1 {\bf a}_1 + n_2 {\bf a}_2 + n_3 {\bf a}_3 + n_4 {\bf b}_1$
spans all of the sphere centers in the tunneled fcc crystal, where the vectors ${\bf a}_i$ 
and ${\bf b}_1$ are defined by (\ref{fcc-a}) and (\ref{fcc-b}), respectively,
the $n_i$ are the integers, and $n_4=0$ or 1. Thus, the corresponding squared
distance from the origin is given by
\begin{equation}
R^2= 3 (n_1^2+ n_2^2 - n_1 n_2 + n_2 n_4) + 2 n_2 n_3 + n_3^2 + n_4^2 + n_3 n_4 - n_1 n_3. 
\label{quad}
\end{equation}
The quadratic form (\ref{quad}) enabled us to determine the theta series 
$\theta_{tfcc}(q)$ for the tunneled fcc crystal up to an arbitrarily
large number of terms (100,000 or more). The first 14 terms of this series are given by
\begin{equation}
\theta_{tfcc}= 1 + 7 q + 4q^2    + 18q^3  +  7q^4   +  16q^5 + 6q^6 + 28 q^7 +4 q^8+
30q^9 +14q^{10} +16 q^{11} +18q^{12} +42 q^{13} + \cdots
\label{theta-tfcc} 
\end{equation}
The theta series is a fundamental characteristic of a packing that encodes
coordination structure information, \cite{Co98} namely, the exponent of $q$ gives
the squared distance of  spheres from a sphere located at the origin
and the associated coefficient is the number of spheres
located at that squared distance. The result (\ref{theta-tfcc}) should
be contrasted with the corresponding theta series for the unvacated fcc packing
given by
\begin{equation}
\theta_{fcc}= 1 + 12q + 6 q^2 + 24q^3  +  12q^4   +  24q^5 + 8q^6 + 48 q^7 + 6q^8+
36q^9 +24q^{10} +24 q^{11} +24q^{12} +72 q^{13} + \cdots
\label{theta-fcc}
\end{equation}
Observe that the coordination shell distances are the same
for both the tunneled fcc crystal and its saturated counterpart,
but the corresponding occupation numbers in the former are always strictly less
than those in the latter. This strict bound is true for every coordination shell.

We note in passing that the corresponding three-dimensional crystals
formed by stacking  Kagom{\' e} layers can be obtained from the 
honeycomb-stacking arrangements by placing the largest nonoverlapping sphere at each
of the midpoints of the bonds joining the spheres in each honeycomb layer (as per the
$d$-dimensional mapping described in the footnote.) The Kagom{\' e}
stackings have a  packing fraction  (i.e., three fourths of $\phi_{\mbox{\scriptsize max}}$) that is considerably higher
than that of the honeycomb stackings, and a contact number of 8.

\subsection{Tunneled HCP Crystal}
   
The alternative layer stacking code (\ref{hcp}) that is associated with the hcp crystal also produces 
a regular tunnel array, exhibiting the same packing fraction (\ref{phi-tunnel}).  However, in this case the 
tunnels have a zig-zag shape with overall orientation parallel to the hexagonal ``c" direction.  
The zig-zag tunnels have three possible lateral directional orientations, depending on whether the
constituent honeycomb layers were stacked periodically as AB, AC, or BC pairs.  
These choices have the zig-zags rotated relative to
one another by plus or minus 60 degrees, when viewed down the hexagonal c axis.
Once again the spheres have seven contacts with immediate neighbors, but those seven neighbors have a non-chiral arrangement. Similarly, in contrast to the
tunneled fcc packing, the hcp counterpart has a unique Voronoi cell,
even if it is substantially less symmetrical than the ones
depicted in Fig. \ref{fcc-voronoi}. The former also possesses 16 edges and 
9 vertices as well as 9 faces: two irregular quadrilaterals, two
pairs of irregular triangles, and three quadrilaterals,
each with a mirror axis of symmetry. Of course, the volume 
of the cell is 3/2 times the volume of an unvacated
hcp packing.

 The tunneled hcp crystal has a fundamental cell containing three spheres.  Again, assuming
that the spheres have unit diameter, the basis vectors locating sphere centers for the fundamental
cell can be designated as follows:
\begin{equation}
{\bf a}_1= \sqrt{3} {\bf i}, \quad  {\bf a}_2= -\frac{\sqrt{3}}{2} {\bf i}  + \frac{3}{2} {\bf j}, \quad
{\bf a}_3= \frac{\sqrt{8}}{3} {\bf k}.
\label{hcp-a}
\end{equation}
The additional two spheres in this fundamental cell are located at
\begin{equation}
{\bf b_1}={\bf j}, \quad {\bf b}_2=-\frac{\sqrt{3}}{6}\,{\bf i}+ \frac{1}{2}\, {\bf j}+
\frac{\sqrt{2}}{3}\, {\bf k}.
\label{hcp-b}
\end{equation}
The vector ${\bf R} = n_1 {\bf a}_1 + n_2 {\bf a}_2 + n_3 {\bf a}_3 + n_4 {\bf b}_1 + n_5 {\bf b}_2$
spans all of the spheres in the tunneled hcp crystal, where the vectors ${\bf a}_i$
and ${\bf b}_i$ are defined by (\ref{hcp-a}) and (\ref{hcp-b}), respectively,
the $n_i$ are the integers, $n_4=0$ or 1, and $n_5=0$ or 1. The corresponding squared
distance from the origin is given by
\begin{equation}
R^2= 3 (n_1^2+ n_2^2 - n_1 n_2 + n_2 n_5) + 2 n_2 n_5 + n_4^2 + n_5^2 + n_4 n_5 - n_1 n_5
+\frac{8}{3}n_3^2 +\frac{8}{3} n_3 n_5.
\end{equation}
The theta series $\theta_{thcp}(q)$ for the tunneled
hcp crystal was also determined up to 10,000 terms or more.
The first 14 terms of this series are
\begin{equation}
\theta_{thcp}= 1 +  7 q + 4q^2    + 2q^{8/3}  +  14q^3   +  6q^{11/3} + 
3q^4 + 8 q^{5} + 12 q^{17/3}+
4q^6 +4q^{19/3} +6 q^{20/3} +14q^{7} +4 q^{22/3} + \cdots
\label{theta-thcp} 
\end{equation}
The result (\ref{theta-thcp}) should be contrasted with the theta series 
for the unvacated hcp packing given by
\begin{equation}
\theta_{hcp}= 1 + 12q + 6 q^2 + 2q^{8/3}  +  18q^3   +  12q^{11/3} + 
6q^4 + 12 q^{5} + 12 q^{17/3}+ 6q^6 +6q^{19/3} + 12q^{20/3} +24q^{7} +6 q^{22/3} + \cdots
\label{theta-hcp}
\end{equation}
Although the coordination shell distances are the same
for both the tunneled hcp crystal and its saturated counterpart,
the corresponding occupation numbers in the former are always less or
equal to those in the latter. The fact that the occupation numbers
in the tunneled hcp and unvacated hcp packings can sometimes
be identical never occurs in the fcc analogs.

\subsection{Tunneled Barlow Packings}

Associated with the infinite number of Barlow packings are the infinite number of tunneled
Barlow packings that are obtained by stacking the honeycomb
layers (two of which are the tunneled fcc and hcp crystals).
However, this infinite set of packings is larger than the unvacated Barlow packings
because, as noted earlier, there are three different choices for the direction of
the tunnels at each stacking stage. Of course, all
of the tunneled Barlow packings have the packing fraction specified
by (\ref{phi-tunnel}). 

At first glance, the tunneled packings may seem to be similar in structure 
to crystal structures involving honeycomb stackings found in nature,
such as hexagonal graphite and boron nitride. \cite{Pa66} However, 
the locations of the honeycomb layers relative to one another
are distinctly different in the latter and, in particular, are
not sublattices of the unvacated Barlow packings.

\subsection{Computer Tests of Jamming Category}

In earlier work, \cite{To01} we suggested that the aforementioned jamming
categories can be tested using numerical algorithms that analyze an equivalent contact 
network of the packing under applied displacements. Subsequently,  a rigorous but
practical algorithm was devised to assess the jamming category
of a sphere packing in this fashion. \cite{Do04,Do04b} The algorithm is based on linear programming 
and is applicable to regular as well as random packings of finite size with 
hard-wall and periodic boundary conditions. If the packing is not jammed, the
algorithm yields representative multi-particle unjamming motions.

We begin by testing the tunneled fcc crystal using this algorithm.
The fundamental (primitive) cell is replicated to form a periodic
unit cell with N spheres, i.e., to form a finite periodic packing. It turns
out that the  tunneled fcc crystal for any finite $N$ under periodic boundary conditions
structures is not collectively jammed. (The actual $N$ used was as large as 500.) 
Although each sphere possesses 7 contacting particles, the 2 contacting particles below and 
above the plane containing the central sphere [cf. (Fig. \ref{fcc-chiral}) can move collectively,
enabling the central sphere to roll into the tunnels. This causes
unjamming of the entire packing. However, if one replaces a single honeycomb layer of the 
tunneled crystal structure at one of the boundaries of the unit cell with a perfect triangular-lattice 
layer of spheres, the aforementioned collective motion is eliminated 
and the packing is strictly jammed. This surface triangular-lattice layer
of particles does not contribute to the density in the infinite-packing limit,
and, therefore, the jamming threshold is given by (\ref{phi-tunnel}).
This reinforcement by a single triangular-lattice layer is also
the reason for the stability of the ball-bearing construction 
depicted in Fig. \ref{fcc-tunnel}. The tunneled hcp crystal packing also has 
an instability without reinforcement, but becomes strictly jammed by inserting a perfect
triangular-lattice layer of particles in the manner described above.

Based on these results, it can be argued that any stacking variant
of the honeycomb layers will also be strictly jammed when reinforced by
a  triangular-lattice layer. Indeed, in computer tests
for random honeycomb stackings with up to 1000 spheres per periodic unit cell,
strict jamming is achieved.

\subsection{Hyperuniformity}

An important characteristic of a packing is the extent to which
long-wavelength density fluctuations are suppressed. A {\it hyperuniform}
point pattern is one in which infinite-wavelength
density fluctuations vanish identically. \cite{To03b} This property
implies that the number variance of sphere centers within
a compact subregion of space (window) grows more slowly than the
volume of the window. All periodic packings are {\it hyperuniform}. \cite{To03b}
However, not all packings that are hyperuniform are
necessarily rigid in the sense of strict jamming, \cite{To03a} especially
if they are not saturated. A packing is {\it saturated} if there is no space
available to add another sphere without overlapping the
existing particles. Thus, both the tunneled fcc and hcp crystals are unusual packings
in that they are hyperuniform (because they are periodic) and strictly jammed
(despite the fact that they are unsaturated). It appears that the
property of hyperuniformity extends to all of the tunneled Barlow packings,
including the purely disordered ones. The reason is that any single honeycomb
layer is itself hyperuniform and translations of the honeycomb
layers in the {\it equally spaced} 
honeycomb planes in the stacked arrangements should not accumulate long-wavelength
density fluctuations. A rigorous proof that the tunneled random stacking variant
is hyperuniform would require one to show that the structure factor vanishes in the limit
of vanishing wavenumber using the methods of Ref. \onlinecite{To03a} that were applied to
other crystal structures.

\section{Discussion and Conclusions}

     Beside their intrinsic relevance for hard-sphere-jamming phenomena, the existence of 
tunneled crystal structures may have substantially broader significance for solid state physics 
and materials science.  For example, the directionality of the parallel arrays of tunnels in the tunneled fcc packing  
(described in Sec. III) might produce some unusual properties that exploit the resulting anisotropy.  
One obvious candidate, say for the linear tunnels in the fcc-parent case, would be separation 
technology for substances whose constituent particles have just the right size to diffuse along 
those tunnels, leaving behind larger impurity particles.  In the event that a metallic element 
or alloy were to be rendered in the form of a tunneled crystal, the electronic characteristics 
would doubtless be strongly influenced by the structural anisotropy. The anisotropic
porosities of a tunneled crystal also suggest that they might
serve as catalytic substances for reactants that fit into these pores. 
Elsewhere, the magnetic properties of the tunneled crystals
are being studied by examining the classical Heisenberg Hamiltionian for Ising, XY,
and Heisenberg spins on these structures. \cite{Bu07}

 It should be kept in mind that the tunneled crystal structures conceivably could be 
synthesized at the atomic scale, or alternatively at a larger length scale they might be 
assembled from spherical colloids. In such instances, the mechanical stability
of the crystals could be enhanced via attractive interactions that are absent in the sphere packings.
In the case of colloids, the attractive interactions can be optimized
for mechanical stability using inverse statistical-mechanical techniques; see Ref. \onlinecite{Re07}
and references therein.

     As stressed earlier, we have no proof that the tunneled-crystal packing fraction 
shown in Eq. (\ref{phi-tunnel}) is the lowest attainable for collectively or strictly jammed sphere 
packings.  It is noteworthy 
that any attempt to remove even a single additional sphere from the tunneled fcc, hcp, 
or hybrid tunneled crystals immediately causes the structure to begin collapsing.  However, 
this observation does not in itself eliminate the possibility of discovering some other 
unrelated class of structures with yet a smaller packing fraction $\phi$.  One might suspect that 
if such a lower density structure were to be created, it might exhibit a number of contacts 
per sphere less than the seven present in all of the cases described herein.  As we
noted earlier, isostatic packings have the minimum 
average number of contacts of six that would be consistent with collective or strict jamming, 
a situation that is usually associated with the amorphous MRJ state. 
This disordered packing, however, has a distribution of contact
numbers that in principle could be as large as the maximum
value of twelve. Ulam conjectured that the maximal
density for packing congruent spheres is smaller than that for any
other convex body (Martin Gardner, private
communication; see also ref. 32). If the tunneled crystals identified in this paper are
indeed the ones that achieve the strict jamming threshold, then it may be
possible that $\phi_{\mbox{\scriptsize min}}$ is itself minimized
by spherical packing elements among all congruent
convex bodies. It is also possible that the tunneled crystals
provide the lowest density strictly jammed structures
that are subpackings of the densest sphere packings, but that true
jamming threshold is achieved by packings that are not
subpackings of the densest packings.

     At least in one, two, and three dimensions, the maximal density
$\phi_{\mbox{\scriptsize max}}$ can only be attained by restricting local 
sphere coordination geometries to a very limited set.  The result is that these maximal-density 
structures exhibit periodic long-range order. This is true even of the
hybrid close-packed crystals, which are only 
partially disordered in their density distributions.  By contrast, the local coordination 
geometries present in amorphous sphere packings are very numerous, but are overwhelmingly 
unsuited as structural elements for attaining $\phi_{\mbox{\scriptsize max}}$.  
We conjecture that an analogous situation 
applies for the jamming threshold $\phi_{\mbox{\scriptsize min}}$.  
The vast majority of amorphous packing coordination geometries are likewise 
unsuited for producing the lowest density collectively or strictly jammed sphere packings.  
Only crystalline structures in three dimensions should be expected to exhibit 
$\phi_{\mbox{\scriptsize min}}$.  Whether the 
tunneled crystals described in this paper are the solution to this minimization remains to be seen.

The authors thank Fiona Burnell and  Shivaji Sondhi for
enlightening discussions concerning the magnetic
properties of the tunneled crystals.
We are also grateful to Yang Jiao for running
the program that tests for jamming on the tunneled
crystals and for creating the bulk of the figures for this paper. This work supported by the National Science Foundation
under Grant No. DMS-0312067.


\begin{thebibliography}{10}
\bibitem{rcp}
As the name implies, the RCP state is traditionally thought of as the highest density
that a large random packing of spheres can attain. Besides the important fact
that randomness was never quantified, it was shown that one can create
packings that are progrsseively denser by decreasing the degree of order,
and hence the concept of the RCP state is ill-defined. See Refs. \onlinecite{To02a}
and \onlinecite{To00} for further details.
\bibitem{order}
Any useful order metric will not assign a value of unity (perfect order)
to all periodic packings, but rather assign a value generally
less than unity  to each according 
its symmetry, vacancy distribution, etc. This has been shown to be the
case in practice. \cite{To00,Ka02b} Therefore, since 
we expect that the jamming-threshold structures are periodic,
point A in Fig. \ref{phi-psi} will likely have a large value of $\psi$
but less than unity.
\end{thebibliography}
\end{document}